# On-Demand Quantum Light Sources for Underwater Communications

**Dominic Scognamiglio[1], Angus Gale[1], Ali Al-Juboori[1], Milos Toth[1,2], Igor Aharonovich[1,2]**
Corresponding Email: igor.aharonovich@uts.edu.au

*1 School of Mathematical and Physical Sciences, Faculty of Science, University of Technology Sydney, Ultimo, New South Wales 2007, Australia*
*2 ARC Centre of Excellence for Transformative Meta-Optical Systems (TMOS), University of Technology Sydney, Ultimo, New South Wales 2007, Australia*

**Abstract**. Quantum communication has been at the forefront of modern research for decades, however it is severely hampered in underwater applications, where the properties of water absorb nearly all useful optical wavelengths and prevent them from propagating more than, in most cases, a few metres. This research reports on-demand quantum light sources, suitable for underwater optical communication. The single photon emitters, which can be engineered using an electron beam, are based on impurities in hexagonal boron nitride. They have a zero phonon line at ~ 436 nm, near the minimum value of water absorption and are shown to suffer negligible transmission and purity loss when travelling through water channels. These emitters are also shown to possess exceptional underwater transmission properties compared to emitters at other optical wavelengths and are utilised in a completely secure Quantum Key Distribution (QKD) experiment with rates of kbits/s.

1. **Introduction**

Fast and secure underwater communications are in ever-increasing demand as research on data transfer and defence technologies evolve[1–3]. Typically, underwater communication is accomplished through longitudinal acoustic waves which propagate through vibrations of atoms4. While they are exceptional over ultra-long range underwater communications[5,6], they are also highly insecure, omni-directional and suffer from very low data rates, especially over these distances[7,8]. For many underwater communication needs such as exploration, research, short range submersibles or military operations[9–11], the separation between the sender and receiver is often a far shorter distance than required for the use of acoustic waves. In these instances communication using standard, transversal electromagnetic waves would be far more beneficial for their increased data transmission rates and high level of security[12,13]. Radio and Infrared waves are commonly used above ground to send and receive information over large distances[14,15], however these methods are impossible underwater due to its absorption properties. To enable the use of electromagnetic waves underwater a different wavelength is needed[16]. Nearly all optical wavelengths of light are absorbed and scattered when propagating through very small bodies of water, yet the use of light in the blue region is able to combat this as the absorption of light in water is at a minimum at ~ 417 nm[16]. To take advantage of this attenuated lasers in the blue and green spectral ranges were also utilised for underwater communications[17,18]. However, while this certainly helps with increasing data rates, photons generated by attenuated lasers have a probabilistic nature[19], which limits their relevance to applications that require maximum information security. To this end, a reliable quantum light source operating in a spectral window suitable for underwater communications is highly sought after, but so far, remains elusive.

In response to this, an exceptional solution has been found to address these problems. Hexagonal boron nitride (hBN) has recently demonstrated the ability to host point defects within the lattice that display single photon emission (SPE)[20–22] Specific defects in hBN emitting at 436 nm are explored as ideal candidates for underwater communications. These defects (termed B-centres) can be deterministically engineered using an electron beam, and they possess extreme photostability[23,24]. Moreover, B-centres are the only single photon emitters known to date, with an emission wavelength near the water absorption minimum (417 nm).

This report aims to use these B-centres in hBN and analyse them through a water collection pathway. It shows that the B-centres transmission and purity are largely unaffected by short bodies of water and are far less affected through the water channels than a similar quantum emitter of a larger wavelength. A demonstration of completely secure Quantum Key Distribution (QKD) is shown as an example of the usefulness of an on-demand, underwater quantum emitter.

2. **Results and Discussion**

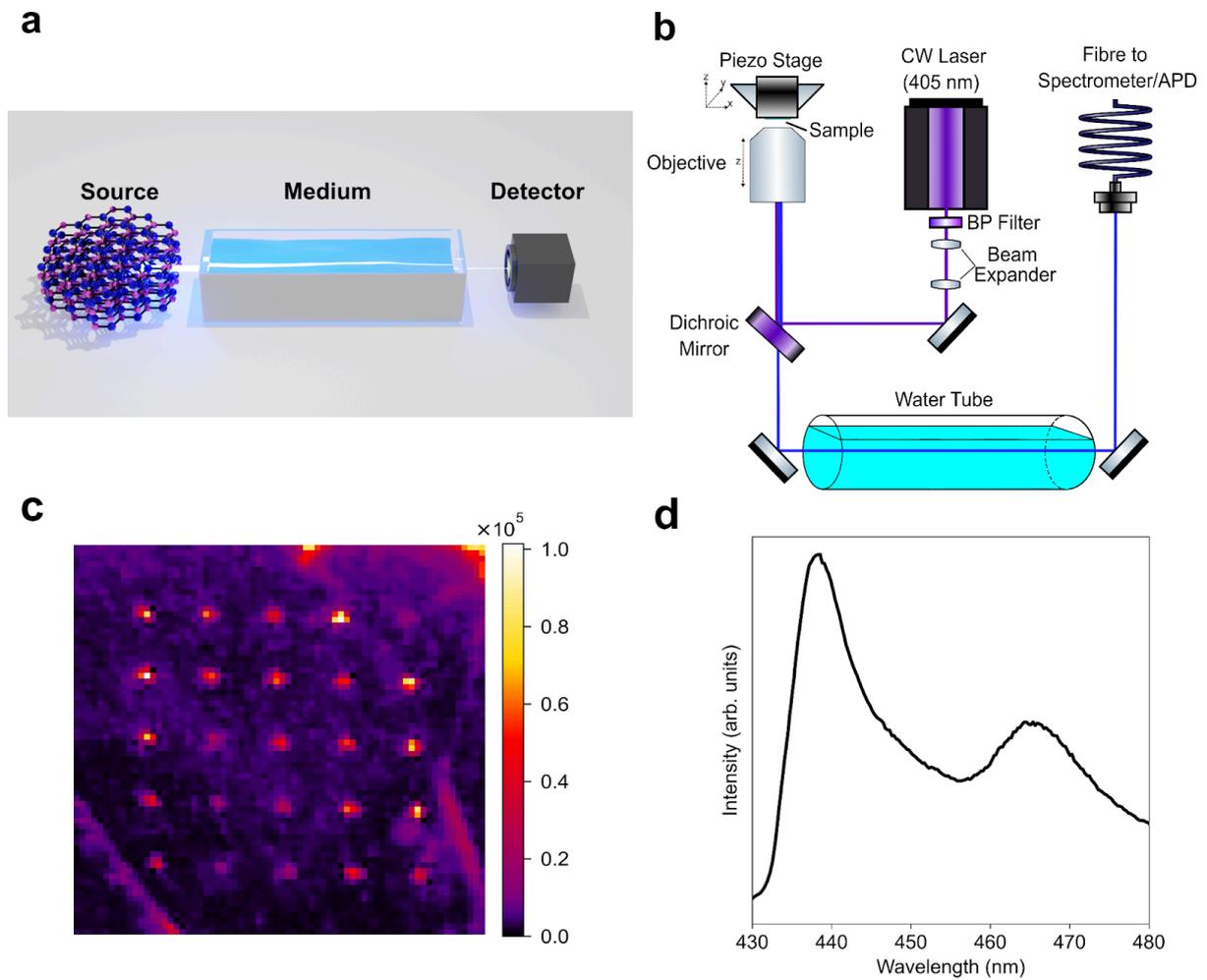

***Fig. 1.*** *(a) 3D representation of a B-centre in hBN propagating through water. (b) Simplified schematic of the confocal setup used in this work, incorporating a 1 m tube of water in the collection path. All components are labelled in the figure. Excitation is performed using a 405 nm laser through a high numerical aperture objective. (c) A photoluminescence (PL) confocal map of a B-Centre array, engineered in a hBN flake. (d) PL spectra of a B-centre ensemble.*

Fig 1(a) depicts the collection of quantum light from B-centres through water using the confocal microscope setup shown in Fig 1(b). It includes a 1 m tube filled with deionised water in the collection path. The tube was sealed on both sides with glass coverslips to allow light transmission with holes drilled into the top and bottom to ensure the tube can be filled with minimal adjustments to the orientation of the tube. Alignment was completed through adjustable mirrors on both sides of the tube. A 5 x 5 array of B-centre ensembles, shown in Fig 1(c), was patterned on an exfoliated hBN flake using an electron beam[23,25]. The difference in intensity between the ensembles is due to the stochastic nature of B-centre creation, with a given electron dose often causing different amounts of B-centres to be formed. A spectrum of one of the bright ensembles from the array is shown in Fig 1(d), recorded through the 1 m water path, illustrating qualitatively the transmission of light emitted by the B-centres.

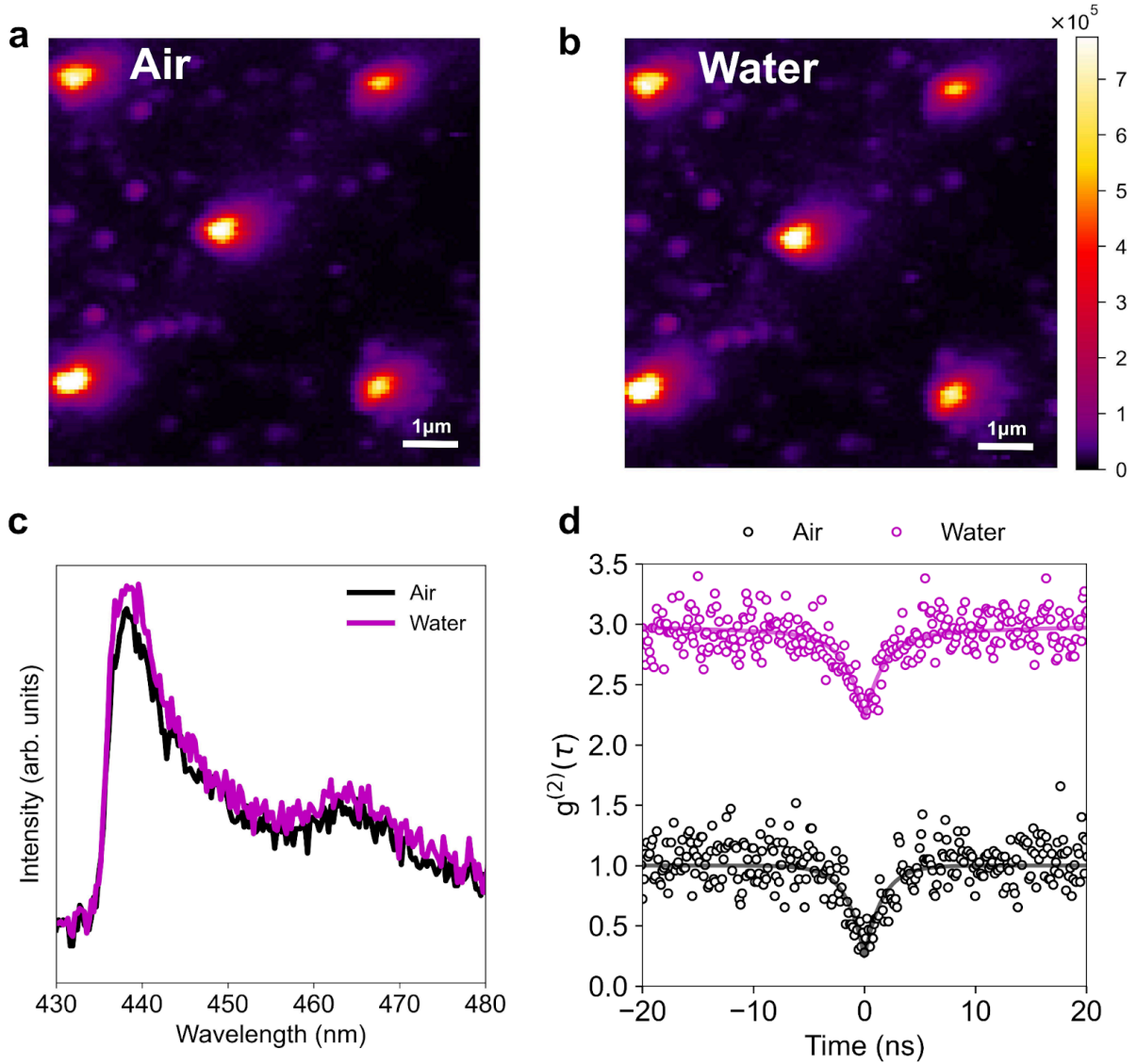

*Fig. 2. Comparison of two PL confocal maps of a small subset of the B-Centre array through both air (a) and water (b). These are seen to be nearly identical (c) Comparison of PL spectra from a single B-Centre collected through air (black) and 1 m of water (magenta). (d) second-order autocorrelation functions of the same SPE collected through air and water, offset for clarity: g(2)(0) = 0.23±0.095 and 0.21±0.097, respectively.*

Higher magnification PL confocal maps were also taken through both air (Fig 2(a)) and water (Fig 2(b)), with a comparison of these maps showing minimal changes in both the large ensemble emitter arrays and smaller isolated SPEs located between the large ensembles. The isolated SPEs are present due to low dose electron exposure by backscattered electrons and electron beam imaging of the hBN flake. Fig 1(c) shows a PL spectrum of a single B-centre under 405 nm excitation with a clear zero phonon line (ZPL) at 436 nm. There is a negligible reduction in the emission intensity, comparing the spectra collected through air (black curve) and 1 m of de-ionized water (magenta curve). The purity of the B-centre was analysed using a Hanbury Brown and Twiss interferometer. Fig 2(d) shows the second-order correlation function with an antibunching value of g(2)(0) ≅ 0.23±0.095 and g(2)(0) ≅ 0.21±0.097, for transmission through air and through water, respectively. These values confirm that single photon purity is maintained after transmission through water. Note that the g(2)(0) values are not background-corrected.

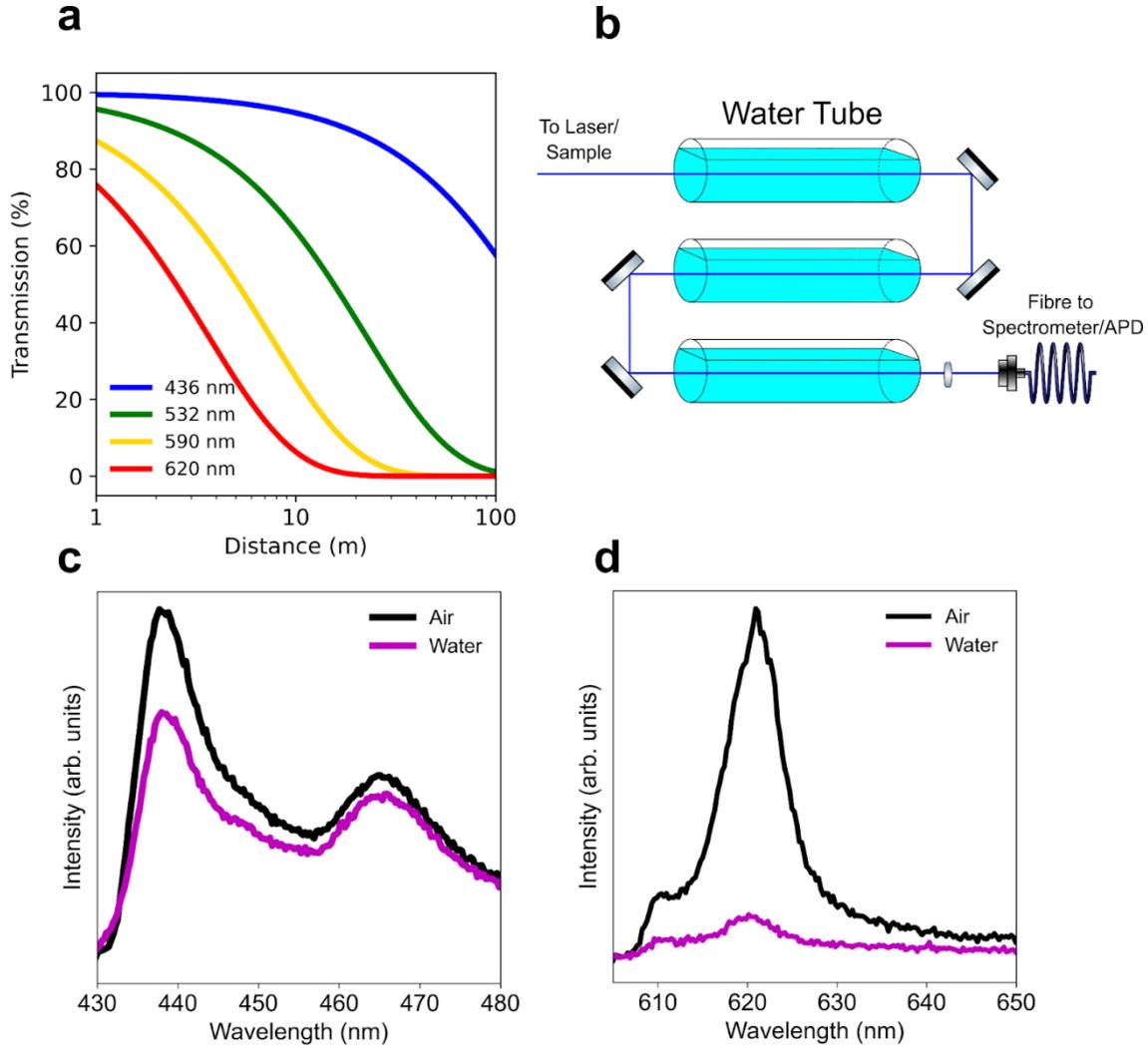

*Fig. 3.* *(a) Optical transmission through water as a function of distance for four selected wavelengths calculated using absorbance coefficients from Pope et al.[16] (b) Schematic of the tube setup used to extend the water path from 1 meter to 3 meters. (c) A comparison of PL spectra of a single B-Centre (ZPL at 436 nm, left panel), and a (d) red SPE (ZPL at 620 nm, right panel), collected through air (black) and 3 m of water (magenta). Absorption is clearly more pronounced in the red spectral range.*

To compare the transmission of the B-centre through water to other optical wavelengths, we employed another quantum emitter in hBN with a ZPL at ~620 nm. Fig. 3(a) shows a plot of light transmission versus distance in water for emission wavelengths of typical SPEs in hBN (using absorption coefficients from Pope et al.[16]). In many publications the absorption coefficient is under debate, with reported absorption coefficients varying across different water types. When comparing all these differing water absorption coefficients it is important to note that most follow the trend of having an extremely low absorption coefficient at ~430[26] nm. While the minimum is still debated, the B-centres engineered by the electron beam in hBN, remain as an ideal candidate for use as a single photon source in quantum key distribution as the emissions. While there are many conflicting reports, Lee et al.[27] reports a lower absorption coefficient of 'pure' seawater at ~ 0.0045 at 436 nm. While calculated distances may vary slightly, 0.0055 m-1 from pope et al. appears to be a conservative coefficient and suitable to use in this case. Over a distance of ~100 meters, transmission losses of only ~40 % are predicted for B-centers compared to losses of >99 % in other spectral ranges. To increase the emission path length through water, two additional 1 m tubes were added to the setup as shown in schematic from Fig 3(b), extending the total pathway through water to 3 m. Fig 3(c) and (d) compares a B-Centre to a typical hBN SPE emitting in the red spectral range (with a ZPL at 620 nm). For the latter, the absorption coefficient in water is ~0.2755 m-1, two orders of magnitude greater than that in the blue spectral range (~0.0055 m-1). To demonstrate this using the two emitters, PL spectra were measured through a longer 3 m path of water. As is clearly seen in Fig 3, the emission in the red spectral range is highly attenuated, compared to that of the B-centre in the blue spectral range. The marginal attenuation of the B-centres are caused by glass reflection along the 3 meter pathway of water that was constructed by aligning three 1 m tubes (due to technical constraints in the laboratory). The interface between air and glass could experience approximately a 4% difference due to reflection, while the interface between air, glass, and water would encounter around a 0.5% difference causing significant transmission losses through the tubes when no water was present. It's crucial to recognize that while the influence of reflections on the glass within a single tube was minimal. When a longer pathway is utilized, such as three tubes as illustrated in Figure 3 (b), these effects become significant, impacting the transmission of the light through the collection path in air (i.e., through the empty tubes with glass covers). Figure 3 (b) and (c) therefore show the 'best' case in air with no obstruction vs the 'worst' case of water obstructing the pathway.. Nevertheless, this result clearly demonstrates the superior performance of the B-centres, which offer a true and unique advantage for real-world underwater communications at room temperature.

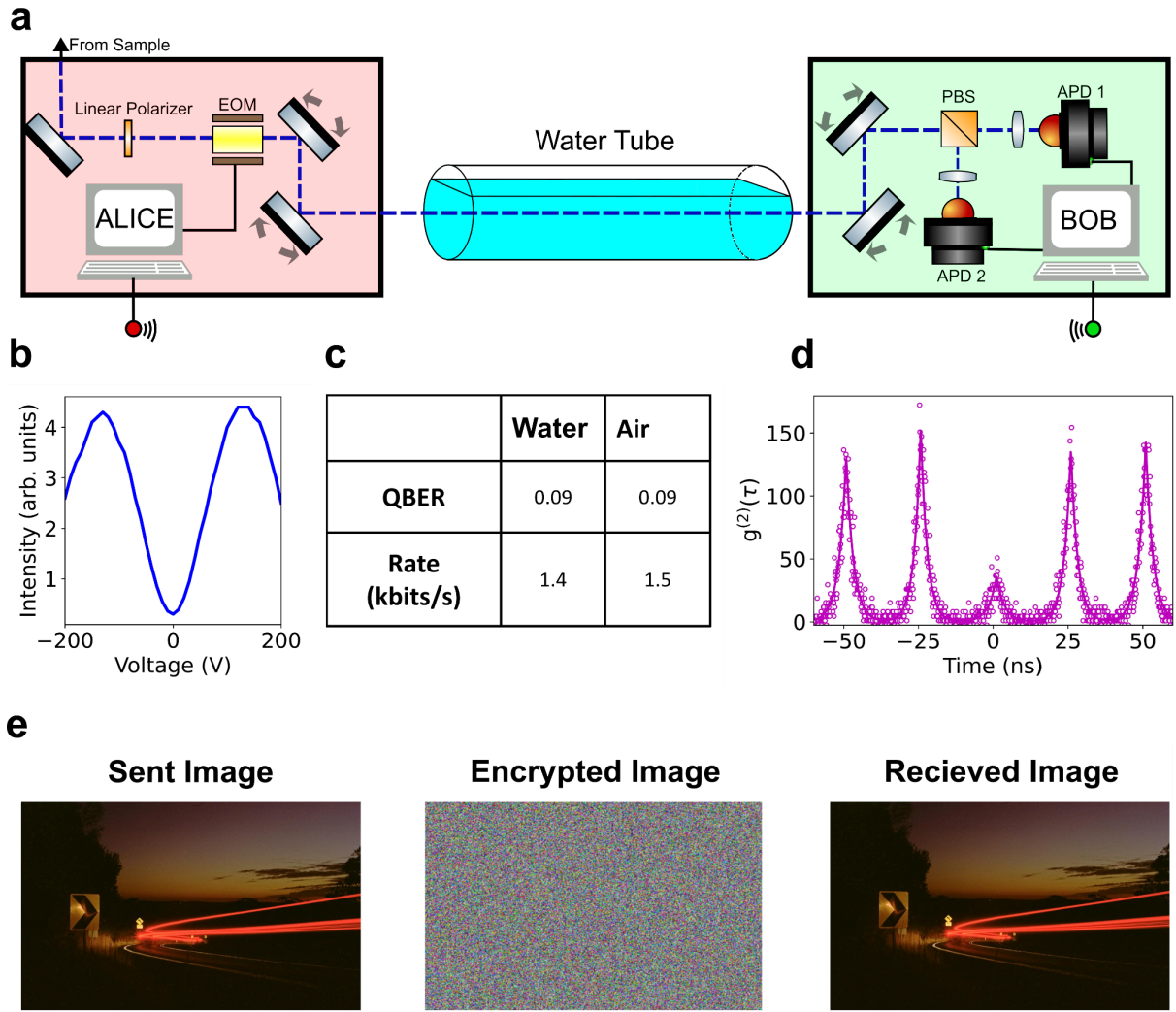

*Fig. 4.* *(a) Schematic of the quantum key distribution (QKD) experiment, with Alice (sender) and Bob (receiver). The quantum-encrypted link, demonstrated using the B-centre, passes through a 1 m water channel, whilst the classical communications are done via a public WiFi channel. EOM - Electro Optical Modulator; PBS - Polarised Beam Splitter. (b) Extinction ratio characterisation of the Electro-Optic Modulator (EOM): Voltage Range vs. Amplitude (c) Table showing a comparison of the data rates and the quantum bit error (%) in air and through water using the same single B-centre. The values in both scenarios are similar. (d) Pulsed second-order autocorrelation function of the B-centre SPE used for underwater QKD. A repetition rate of 30 MHz was used resulting in g(2)(0) = 0.27 (e) Images representing the QKD process.*

To prove the utility of the B-centres, a demonstration of QKD through water was then undertaken using a simplified version of BB84[28,29], which utilizes one basis (rectilinear) and two polarization states (horizontal and vertical) instead of four in full BB84. In our setup, Alice generates a sequence of random integer numbers (0 and 1) which serve as bits 0 and 1 respectively, and associates 0 to a certain voltage that when applied to the Electro-optic Modulator (EOM), induces a horizontal polarization and associates 1 with the value that induces vertical polarization. Fig 4.(a) shows the schematic of the setup used for QKD. Alice (Sender) sends a string of randomly polarized photons to Bob through the quantum channel (water in this case). On Bob's end, horizontally-polarized photons will pass straight through polarizing beam splitter (PBS), and be detected at APD1 and get registered as bit (0), while vertically-polarized photons will be reflected by the PBS and get detected by APD2 and get registered as bit (1). Since Alice and Bob are always using the rectilinear basis, every bit Bob records corresponds directly to a bit Alice sent. They publicly discuss which photons were successfully received and measured by Bob without revealing their values. Any photons not detected by Bob (due to loss in the channel or other reasons) are discarded from Alice's string. The remaining bits in both Alice and Bob's strings form the raw key. Next step is estimating the quantum bit error rate (QBER), which indicates the fraction of bits that are in error in the raw key generated after the sifting process. It is calculated by comparing a subset of the raw keys of Alice and Bob using: QBER = number of erroneous bits/total number the bits exchanged. The exchanged subset of the raw keys (50% in this experiment) used for the QBER estimation is then discarded. The QBER estimation is followed by the reconciliation process (error correction) to ensure that Alice and Bob end up with identical secret keys. Using the estimated QBER, Alice and Bob employ classical error correction algorithms to correct the discrepancies in their raw keys. There are many reconciliation protocols, each of which has its own strengths and weaknesses. For our experiment, the Cascade protocol was used, which is an iterative error correction protocol. It involves multiple passes over the data, where in each pass, the data is divided into blocks, and parity checks are performed to identify and correct errors. To eliminate any potential data leakage that might have happened during the reconciliation process, both Alice and Bob apply a process called Privacy Amplification, which uses universal hash functions to shorten the key in a way that any remaining information an eavesdropper might have becomes negligibly small.

The measured QBER for this experiment was 0.09. This value is limited by the extinction ratio of the EOM. The EOM's critical role in our setup is to induce four distinct polarization states, fundamental to the BB84 encoding scheme. However, the calculated extinction ratio, determined by dividing the lowest by the highest amplitude values obtained across the EOM's voltage range, revealed a value of 0.09, as shown in Figure 4 (b). This figure underlines the modulator's limited effectiveness in generating high-contrast, distinguishable polarization states. The consequence of this limited extinction ratio is a pronounced overlap in the polarization states, leading to increased error rates in state discrimination during the key exchange process. This overlap consequently elevates the QBER to 0.09, underscoring the direct correlation between the modulator's physical limitations and the performance of the QKD system. Note, however, QKD is considered to be completely secure at QBER >0.11[30]. Therefore this measurement successfully demonstrated that water has a negligible effect on the collection and encryption of light when using wavelengths near the minimum absorption coefficient of water. The obtained rates measured in Fig 4 (c) are the amount of photons detected by Bob's APDs per second after performing the sifting process. This diminishes the rates counted by Bob as 50% of counts are sacrificed to generate the QBER. The triggered second-order autocorrelation function of the B-centre used in the QKD demonstration and collected through 1 m of water is seen in Fig 4(d). The $g2(0)$ of ~0.3 is relatively high for use in a QKD experiment but there have been multiple instances of higher purity optical emitters, including B-centers, in hBN[22–24,31].

Fig 4 (e) demonstrates the QKD process. The image of the road exists on Alice's device shown on the left. The key from the QKD experiment is then used as a one time key pad to encrypt the image. The image is then classically transmitted to Bob, who uses the secure key generated on their end to decrypt and view the image on the left. For use over long distances, the underwater rate of 1500 bits per second was also heavily limited by the performance of the EOM when used in the blue spectral region, with an ~ 80 % transmission loss measured at 488 nm. The avalanche photodiode used for the experiment (Excelitas SPCM-AQRH) has poor quantum efficiency (~15%) at these wavelengths. This gave us a collection efficiency of roughly 10-3, while this is low, there are many ways this collection efficiency can be improved. Additional cleaning measures and the use of high quality hBN to minimise background fluorescence are possible steps toward achieving higher purity and rates. Beyond optimisation of the setup, a further improvement in rates can be achieved using purpose-built avalanche photodiodes with enhanced efficiency in the blue spectral range, as well as engineering of cavities and waveguides to enhance the emission rate of the B-centre[32–34]. In principle, with typical lifetimes of a few ns, and minimal losses in water, rates of GHz can be achieved, resulting in QKD rates in excess of Mbp/s over a few km, outperforming current acoustic wave data rates and high rates achieved using an attenuated green laser source.Alternate technologies are in circulation in the realm of optical underwater technologies. Classical cryptography is also a viable alternate for secure underwater communications. However, the emergence of large-scale quantum computers poses a future threat to all classical cryptographic systems, underscoring the importance of developing quantum-resistant encryption methods. While the authors agree that practical, real world QKD (particularity underwater in this case) faces many obstacles and challenges, one cannot completely ignore the potential benefits that the technology would bring. Any optical technology developed for use underwater also must also acknowledge extra background light scattering into the optical channel. Whilst light propagating underwater is certainly affected by background light and scattering, sophisticated filtering methods for underwater optical systems are already in development[35,36]. With technology in this space continuing to advance, filtering of background light will only improve.

To conclude, we demonstrated robust, on-demand single photon emission in the blue spectral range, suitable for communications through water. We have further demonstrated secure underwater QKD, hence introducing a promising avenue for future employment of hBN defects in underwater communications.

3.  **Methods**

Photoluminescence Measurements: the confocal microscope shown in fig 2(a) uses a 405 nm continuous-wave (CW) laser (PiL040X, A.L.S. GmbH) with a clean-up filter which is incident upon the sample through a 100x Nikon objective (0.9 NA). The emitted light is then collected and transmitted through a dichroic mirror, and directed through the body of water used. This project used 1m PVC pipes purchased from Bunnings with glass cover slides epoxied on each side to seal the water in. The top and bottom were screwed with a funnel used to fill the pipe with water. Deionized water was used. The light was then collected into a multimode fibre and directed into either a Hanbury Brown and Twiss (HBT) interferometer setup[37] or spectrometer (Princeton Instruments, Inc.). When performing the QKD experiments a separate QKD setup was used, which is outlined in the main body of the text due to its complexity.

Sample Preparation: hBN flakes were mechanically exfoliated with scotch tape onto Si/SiO2 substrates (285 nm oxide layer). Prior to exfoliation, substrates were cleaned by 30 min sonication in acetone, isopropanol and dried using nitrogen. The surfaces were further cleaned by exposing the substrates to a commercial UV ozone bath (ProCleaner Plus, Bioforce Nanosciences Inc.) for 10 minutes. Post exfoliation, substrates with flakes were annealed on a hotplate at 500 °C for 1 hour.

Electron Beam Irradiation: A Thermo Fisher Scientific Helios G4 Dual Beam microscope was used to create a 5 x 5 array of B-centre ensembles with 5 μm spacing using inbuilt patterning software. The electron beam had a current of 0.8 nA and beam energy of 5 keV. These parameters were chosen based on literature22. Individual B-centres were created between the arrays due to additional SEM imaging as seen in the PL confocal map in Fig 2(b).

4. **Funding**

Australian Research Council (CE200100010, FT220100053), Office of Naval Research Global (N62909-22-1-2028)

5. **Acknowledgements**

The authors thank the ANFF node of UTS for access to facilities. The authors also thank Minh Nguyen and Mehran Kianinia for their assistance with the experimental setups.